\documentclass[prl,aps,showpacs,twocolumn,floatfix]{revtex4}
\usepackage{epsf}
\usepackage[dvips]{color}
\usepackage[nosumlimits]{amsmath}
\newcommand {\be}{\begin{equation}}
\newcommand {\ee}{\end{equation}}
\newcommand {\bea}{\begin{eqnarray}}
\newcommand {\eea}{\end{eqnarray}}

\begin{document}

\title{Vortices in attractive Bose-Einstein condensates in two dimensions}
\author{L.~D. Carr$^{1,2}$ and Charles W. Clark$^2$\\}
\affiliation{1. Physics Department, Colorado School of Mines, Golden, CO 80401\\
2. Electron and Optical Physics Division, National Institute of
Standards and Technology, Technology Administration, U.S.
Department of Commerce, Gaithersburg, Maryland 20899}

\begin{abstract}
The form and stability of quantum vortices in Bose-Einstein
condensates with attractive atomic interactions is elucidated. They
appear as ring bright solitons, and are a generalization of the
Townes soliton to nonzero winding number $m$.   An infinite sequence
of radially excited stationary states appear for each value of $m$,
which are characterized by concentric matter-wave rings separated by
nodes, in contrast to repulsive condensates, where no such set of
states exists. It is shown that robustly stable as well as unstable
regimes may be achieved in confined geometries, thereby suggesting
that vortices and their radial excited states can be observed in
experiments on attractive condensates in two dimensions.
\end{abstract}


\maketitle

The study of vortices has a long and illustrious scientific history
reaching back to Helmholtz and Lord Kelvin in the nineteenth
century~\cite{saffman1992}. Vortices associated with quantized
circulation are a central feature of
superfluidity~\cite{donnelly1991}.
Singly-~\cite{matthews1999,madison2000} and
multiply-quantized~\cite{shin2004b} vortices have been observed in
Bose-Einstein condensates (BECs) with repulsive atomic interactions.
Complex vortex structures have been shown to be stable in repulsive
BEC's, including vortex dipoles~\cite{crasovan2003,qizhou2004} and
vortex rings~\cite{anderson2001}.  The nonlinear Schr\"odinger
equation (NLSE), which provides an excellent description of BECs at
the mean-field level~\cite{dalfovo1999}, supports vortex solutions,
which have been studied extensively in the case of repulsive atomic
interactions~\cite{fetter2001}.  The main goal of this Letter is to
clarify the meaning and nature of single vortices and their excited
states in BECs with {\it attractive} interactions, and thus
encourage experimental exploration of stable and unstable two
dimensional (2D) BECs.

Solutions to the NLSE with attractive, or focusing nonlinearity,
in contrast to repulsive, or defocusing nonlinearity, are unstable
in free space in three dimensions (3D) and are stable in one
dimension (1D)~\cite{sulem1999}.  The imposition of an external
potential in 3D can produce a long-lived metastable
regime~\cite{ruprecht1995,ueda1998}. In the metastable and
unstable regimes, growth and collapse cycles~\cite{sackett1998}
and implosion~\cite{donley2001} have been studied in 3D. In the
stable regime, bright soliton propagation~\cite{carr2002b} and
interactions~\cite{strecker2002} in a waveguide have been
investigated in 1D~\cite{caveat}.

\begin{figure}[t]
\epsfxsize=8cm \epsfysize=5.8cm \leavevmode
\epsfbox{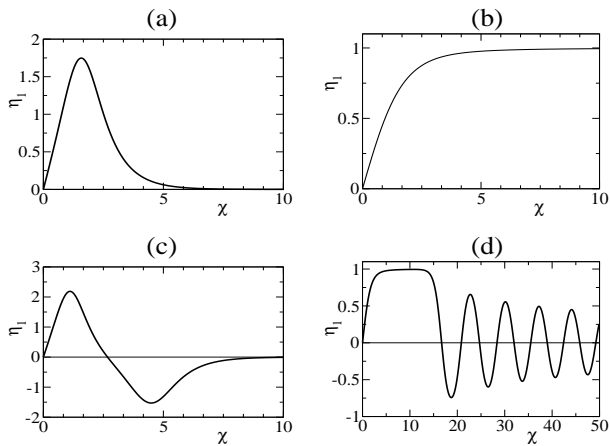} \caption{\label{fig:1} {\it  A quantum
vortex of winding number $m=1$}: (a) attractive case; (b) repulsive
case.  A radially excited state: (c) the first excited state in the
attractive case; (d) in the repulsive case, a radially excited state
requires an infinite number of nodes and asymptotically resembles
the Coulomb function~\cite{carr2004i}. The radial dependence of the
order parameter of an infinitely extended condensate is depicted.
Note that all axes are dimensionless.}
\end{figure}

The critical dimensionality for the NLSE is
2D~\cite{sulem1999,fischer2004}.  We will show that quantum vortices
and their radially excited states can be made robustly stable in
confined, attractive 2D condensates and are a generalization of the
Townes soliton~\cite{chiao1964} to nonzero winding number $m$. The
Townes soliton is fundamental to understanding the self-similar
collapse of solutions to the 2D NLSE~\cite{moll2003}. Its
generalization to winding number $|m|\geq 1$ has been studied in the
context of optics, where such solutions are called ``ring-profile
solitary waves'' or ``spinning bright solitons''~\cite{firth1997}.
This is in fact the attractive analog of the well-known single
vortex solution in repulsive condensates~\cite{donnelly1991}, as we
will show.  An example contrasting the form of vortices in
attractive and repulsive BECs is illustrated in Fig.~\ref{fig:1}(a)
and~\ref{fig:1}(b).

In previous studies of quantum vortices in condensed
matter~\cite{donnelly1991} and optical systems~\cite{firth1997} with
attractive nonlinearity, the phase variation derived wholly from
circulation of matter about the central vortex core.  In this work,
we investigate the most general type of single-vortex stationary
solutions in attractive BECs for which the phase of the order
parameter also alternates sign along radial lines.  For each winding
number $m$, we find a denumerably infinite set of radially excited
states characterized by the successive formation of nodes at
$r=\infty$.  An example of such an excited state is illustrated in
Fig.~\ref{fig:1}(c). In contrast, Fig.~\ref{fig:1}(d) shows how
extremely different is the repulsive case.

Consider an order parameter of the form \be
\psi(r,\phi,t)=f_m(r)\exp(i m \phi)\exp(-i \mu
t/\hbar)\exp(i\theta_0)\,, \label{eqn:assume}\ee which solves the 2D
NLSE \be [-(\hbar^2/2M)\nabla_{r,\phi}^2 + g_{\mathrm{2D}}
|\psi|^2+V(r)]\psi= i\hbar\partial_t \psi \,,
\label{eqn:gpe2dunits}\ee with $V(r)$ a central potential in two
dimensions, $m$ the winding number, $M$ the atomic mass,
$g_{\mathrm{2D}}\equiv (4\pi\hbar^2
a_s/M)\sqrt{M\omega_z/2\pi\hbar}$ the 2D atomic interaction
strength, $a_s<0$ the $s$-wave scattering length, and $\mu$ the
chemical potential or eigenvalue. In Eq.~(\ref{eqn:gpe2dunits}) it
was assumed that the BEC remains in the ground state in a harmonic
trap of angular frequency $\omega_z$ in the $z$
direction~\cite{petrov2000}, so that $r\equiv\sqrt{x^2+y^2}$. Then,
taking $\theta_0=0$ and defining
$\eta_m(\chi)\equiv\sqrt{|g_{\mathrm{2D}}/\mu|} f_m(r)$,
$\chi\equiv(\sqrt{2M|\mu|}/\hbar)r$, one obtains a rescaled 2D NLSE
of form \be \eta_m''+\chi^{-1}\eta_m'-m^2\chi^{-2}\eta_m + \eta_m^3
-V(\chi)\eta_m +\sigma_{\mu}\eta_m=0 \, ,\label{eqn:gpe2d}\ee where
$\sigma_{\mu}=\mathrm{sgn}(\mu)=\pm 1$. The solutions to this
nonlinear second order ordinary differential equation describe
quantum vortices and their radially excited states in an attractive
BEC in an external potential $V(\chi)$. Examples for
$\sigma_{\mu}=-1$ and $V=0$ are shown in Fig.~\ref{fig:1}(a)
and~\ref{fig:1}(c).

The form of the radial wavefunction $\eta_m$ can be obtained from
Eq.~(\ref{eqn:gpe2d}) by numerical shooting
methods~\cite{press1993}.  This requires the initial conditions
$\eta_m(\chi_0)$, $\eta_m'(\chi_0)$, which can be obtained via a
power series around $\chi=0$: \be \eta_m(\chi)=\sum_{j=0}^{\infty}
a_j \chi^{2j+m} \, ,\label{eqn:taylor} \ee where the $a_j$ are
coefficients.
Note that $\eta_m(\chi)\rightarrow a_0\chi^m$ as $\chi\rightarrow
0$. Upon substitution into Eq.~(\ref{eqn:gpe2d}) and solution of the
resulting simultaneous equations, one finds that all coefficients
$a_j$ for $j\neq 0$ can be expressed as a polynomial in powers of
$a_0$.
Thus the power series shows
that $a_0$, the coefficient of the first nonzero term in the
series, together with the winding number $m$, is sufficient to
determine the entire solution.  Note that we take $a_0\geq 0$ and
$m\geq 0$; solutions identical in form exist for $a_0<0$ and/or
$m\leq 0$.

\begin{figure}[t]
\epsfxsize=8cm \epsfysize=6.0 cm \leavevmode
\epsfbox{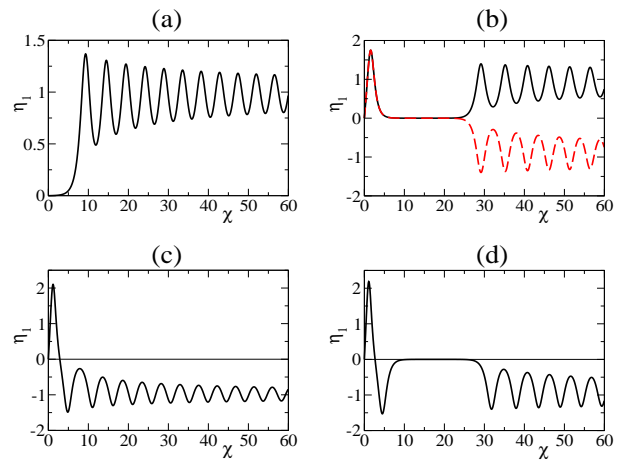} \caption{\label{fig:2} {\it The
development of excited states of a quantum vortex in an attractive
BEC.} (a) For small $a_0$, no quantum vortex is formed. (b) As
$a_0\rightarrow a_0^{\mathrm{vortex}}$ from below (solid black
curve) or above (dashed red curve), the oscillations are pushed out
towards $\chi=\infty$. For $a_0 = a_0^{\mathrm{vortex}}$ a node
appears at $\chi=\infty$, and a true vortex appears. (c) For still
larger $a_0$, the node moves inwards.  (d) As $a_0$ approaches a
second critical value, the oscillations are again pushed out, until
a second node appears at $\chi=\infty$.  In this way one obtains the
first excited state. An infinite sequence of excited states can be
constructed in this way.  Note that the case of an infinitely
extended condensate of winding number $m=1$ is depicted and all axes
are dimensionless.}
\end{figure}

We first consider the case of no external potential, $V(\chi)=0$. By
following the form of the wavefunction as a function of $a_0$, one
observes its entire development.  The case of $m=1$ and
$\sigma_{\mu}=-1$ is illustrated in Fig.~\ref{fig:2}. For very small
$a_0$, the linear, divergent Bessel function solution $K_m(\chi)$ is
recovered for small $\eta_m(\chi)$. However, the nonlinearity
regulates the divergence as $\eta_m$ becomes on the order of unity.
It subsequently oscillates around unity, with the oscillations
damping away as $\chi\rightarrow\infty$ (Fig.~\ref{fig:2}(a)).  As
$a_0$ approaches a critical value $a_0^{\mathrm{vortex}}$, the
oscillations are pushed out towards $\chi=\infty$ and a localized
central peak appears in the wavefunction near the origin
(Fig.~\ref{fig:2}(b)). For $a_0=a_0^{\mathrm{vortex}}$
precisely~\cite{carr2004i}, a node forms at $\chi=\infty$ and a
quantum vortex is obtained.  As $a_0$ is increased, the node moves
inwards and oscillations resume beyond it (Fig.~\ref{fig:2}(a)). To
move the oscillations towards infinity again, one must increase
$a_0$ towards a second critical value (Fig.~\ref{fig:2}(d)).  When
this value is reached precisely, one obtains the first excited
state, with no oscillations past the first node, and a second node
appears at $\chi=\infty$.  In this way one can construct an infinite
set of excited states by increasing $a_0$ to successive critical
points. These critical points are always characterized by the
formation of an additional node at $\chi=\infty$.

The appearance of a denumerably infinite set of critical points is
in strong contrast to the case of a vortex in a repulsive
condensate. As we have shown in previous work~\cite{carr2004i}, for
a repulsive BEC there is only {\it one} critical value of $a_0$,
i.e., $a_0^{\mathrm{vortex}}$. Larger values of $a_0$ result in an
infinite number of nodes and the wavefunction asymptotically
resembles the Coulomb function, as illustrated in
Fig.~\ref{fig:1}(d).  We called these {\it ring solutions}, in
contrast to vortex solutions, as they have a different analytic
structure and asymptotic behavior.  Thus, in an infinitely extended
system, vortices in repulsive condensates cannot be radially excited
in a stationary way, whereas in attractive condensates they have a
denumerably infinite set of excited states.  We note that, for
positive chemical potential, i.e., $\sigma_{\mu}=+1$, one can also
find ring solutions in attractive condensates.

For winding number $m=0$ and $a_0=2.20620086\ldots$, one obtains the
Townes soliton.  Increasing $a_0$ results in the formation of
successive nodes at $\chi=\infty$, just as for $m=1$, i.e., one
finds the radially excited states of the Townes soliton.  In this
special case, it has been formally proven that an infinite set of
radially excited states corresponding to the formation of nodes
exists, and that the Townes soliton is the unique ``ground state''
in that sequence~\cite{sulem1999}.   It is in this sense that the
vortex solutions we have described are generalizations of the Townes
soliton to nonzero winding number.  All such solutions wherein a
node has formed at infinity are normalizable.

In order to study attractive BECs in experiments, it is vital to
consider stability properties in confined systems.  The special
stability properties of 2D can be illustrated by a simple
variational study. Consider the variational ansatz \be
\psi(r,\phi,t) = A\, r^m e^{-r^2/2r_0^2}e^{i m \phi}e^{-i\mu
t/\hbar}\,,\label{eqn:ansatz}\ee where $r_0$ and $A$ are variational
parameters, subject to the power law potential $V(r)=V_0 r^j$, $j >
0$. Integrating Eq.~(\ref{eqn:gpe2dunits}) over $\psi^*(r,\phi,t)$,
 one finds simultaneous equations for the chemical potential and
$g_\mathrm{2D}$.  Then, using the normalization $\int d^2r
|\psi|^2=N$ to eliminate $A$, where $N$ is the total number of
atoms, one obtains $\mu(\mathcal{N})$ parametrically in $r_0$:
\begin{eqnarray}
\mu=\frac{-\hbar^2\Gamma(m+2)} {2 M r_0^2\Gamma(m+2)}
+\frac{V_0(1+j)\Gamma(m+\frac{j}{2}+1)\,r_0^{j}}
{\Gamma(m+2)}\,,\\
\mathcal{N}=\frac{\Gamma(m+2)}
{\pi^{-\frac{1}{2}}\Gamma(m+\frac{1}{2})} - \frac{MV_0 j
\Gamma(m+\frac{j}{2}+1)\,r_0^{j+2}}
{\pi^{-\frac{1}{2}}\hbar^2\Gamma(m+\frac{1}{2})}\,,\end{eqnarray}
where $\mathcal{N}\equiv MN|g_{\mathrm{2D}}|/2\pi\hbar^2=2|a_s|N M
\omega_z/\sqrt{2\pi}\hbar$ is the dimensionless nonlinearity. The
solution is radially stable when the Vakhitov-Kolokolov (VK)
criterion~\cite{vakhitov1973} $d\mu/d\mathcal{N} \leq 0$ is
satisfied.  One finds that the VK criterion always holds for
$\mathcal{N} <\mathcal{N}_c$, where \be
 \mathcal{N}_c=2\sqrt{\pi}\,\Gamma(m+2)/\Gamma(m+1/2)
\label{eqn:radialstab}\,.\ee Equation~(\ref{eqn:radialstab}) is {\it
independent} of both $V_0$ and $j$, i.e., radial stability does not
depend on the details of any positive power law potential. In the
limit in which $j\rightarrow\infty$, one obtains the same result for
a cylindrical hard-wall potential. When $\mathcal{N}\geq
\mathcal{N}_c$, the total energy $E[\psi]=
N[\mu-(g_{\mathrm{2D}}/2)\int d^2r|\psi|^4]\rightarrow -\infty$ as
$r_0\rightarrow 0$, meaning that the wavefunction is unstable and
implodes.  When $\mathcal{N}< \mathcal{N}_c$, the energy has a {\it
global} minimum at some finite $r_0$.  In contrast, in 3D, the
minimum, when it exists, is always local, so that, at best, one
obtains metastability~\cite{perez1997,ueda1998}.

\begin{figure}[t]
\epsfxsize=8cm \epsfysize=6cm \leavevmode
\epsfbox{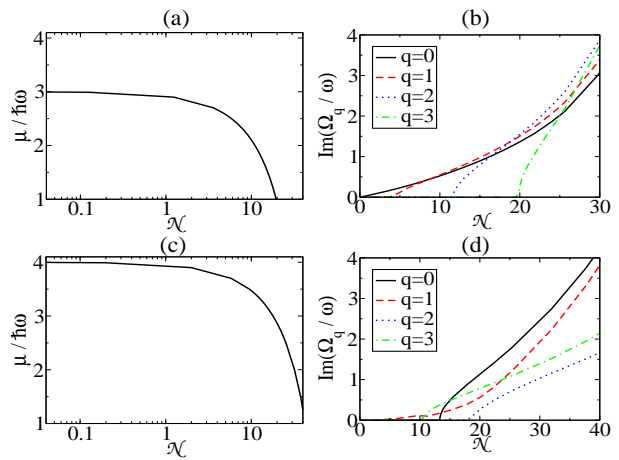} \caption{\label{fig:3} (color online)
{\it Stable regimes of quantum vortices in confined attractive
BECs.} Shown are the chemical potential spectra and linear
instability frequencies as a function of the nonlinearity for
(a)-(b) the first excited state of the Townes soliton, and (c)-(d)
the first excited state of a singly quantized vortex $m=1$, all in a
harmonic potential. In (b),(d) the winding number of the instability
mode is denoted by $q$ and the instability time given by
$2\pi/\mathrm{Im}(\Omega_q)$. Note that all axes are scaled to
harmonic oscillator units.}
\end{figure}

We now consider in detail the most experimentally relevant case,
i.e., $j=2$ and $V_0=M\omega^2/2$, a harmonic trap.  Note that
typical trapping frequencies are on the order of $\omega=2\pi \times
100$ Hertz, which gives a time scale of $T\equiv 2\pi/\omega = 10$
ms.  For no external potential, $V(r)=0$, it is well known that the
Townes soliton ($m=0$) is radially unstable, while the quantum
vortex of winding number $m=1$ (Fig.~\ref{fig:1}(a)) is azimuthally
unstable. The variational analysis of
Eqs.~(\ref{eqn:ansatz})-(\ref{eqn:radialstab}) suggests that the
addition of a confining potential stabilizes the solutions radially
for general $m$.  However, this simple variational study did not
allow for azimuthal stability. Both radial and azimuthal stability
can be determined by linear stability analysis, i.e., the
Boguliubov-de Gennes equations (BDGE)~\cite{fetter2001}, a standard
method which we do not reproduce here, for the sake of brevity.  The
case of general $m$ \textit{without} radial excitations has been
studied previously~\cite{dodd1996a,pu1999,saito2002}.  It was found
that for sufficiently small $\mathcal{N}$ the vortex with $m=1$ is
stable, while for $m\geq 2$ the solution is unstable to quadrupole
oscillations, though instability times can be much greater than $T$.
We can therefore say that for small $\mathcal{N}$ vortices of
winding number $m\geq 2$ are \textit{experimentally stable}.

Applying the BDGE's, we studied the azimuthal and radial stability
of the radially excited states of the Townes soliton. The results
for the first excited state, which has a matter-wave ring separated
from a soliton core by a radial node, are illustrated in
Fig.~\ref{fig:3}(a)-(b). The winding number of each Boguliubov mode
is denoted by $q$. The frequency of each mode is denoted by
$\Omega_q$ with instability time $t_q\equiv
2\pi/\mathrm{Im}(\Omega_q)$. From Fig.~\ref{fig:3}(b), it is
apparent that a radial instability ($q=0$) occurs for all
$\mathcal{N}$, since $\mathrm{Im}(\Omega_0)\neq 0$.  However, one
finds experimental stability for small $\mathcal{N}$, since
$t_q/T\gg 1$. Other modes become unstable at higher $\mathcal{N}$,
starting with the dipole, $q=1$.  For large $\mathcal{N}$ the
dominant instability occurs in the quadrupole mode.

In Fig.~\ref{fig:3}(c)-(d) is shown the same stability analysis for
the first excited state of the $m=1$ vortex, which has two
concentric matter-wave rings separated by a radial node.  Unlike the
excited states of the Townes soliton, here the solution is linearly
stable for $\mathcal < 5.5$.  The solutions continues to be
experimentally stable up to $\mathcal \sim 10$, although a dipole
instability occurs for $t_1 \gg T$.  For large $\mathcal{N}$ the
dominant instability is radial.  For the radially excited states of
both the Townes soliton and the $m=1$ vortex, the VK criterion
fails, as is apparent in Figs.~\ref{fig:3}(a) and~\ref{fig:3}(c).
This is first instance we know of where this criterion, which so far
as we know has never been proven formally, fails.  For winding
number $m=2$ we find that the first excited state is linearly most
unstable to quadrupolar excitations, although for
$\mathcal{N}\lesssim 10$ it is experimentally stable.

In experiments, one may access the stable regime of quantum vortices
and their excited states via a Feshbach resonance, which allows for
extremely small scattering lengths on the order of the Bohr radius;
this technique was used to successfully create bright
solitons~\cite{carr2002b,strecker2002}.  Vortices may be created by
rotating the condensate; alternatively, one may start with a single
vortex in a repulsive condensate and adiabatically switch the
scattering length from positive to negative.  Excited states may be
made by phase imprinting through a pinhole mask, by passing a
tightly focused laser pulse through the condensate center, or by
utilizing a doughnut mode of a laser. It seems likely that the most
useful approach will use these methods to first create dark ring
solitons~\cite{theocharis2003} on a repulsive condensate with a
central vortex~\cite{carr2004i}, and then tuning the scattering
length to be negative.  An important point in the assumption of the
2D regime is that, in order for our stability criterion to be valid,
the condensate must remain oblate, so that excitations in $z$ do not
occur; this may be achieved by a correct choice of experimental
parameters.

In conclusion, we have shown that quantum vortices and their
radially excited states in attractive BECs can be created stably in
confined systems.  We contrasted vortices in attractive BECs, which
can be thought of as ring bright solitons or spinning Townes
solitons, to their counterparts in repulsive BECs. We showed that
there exists a denumerably infinite set of excited states which, in
an infinitely extended condensate, correspond to the creation of
nodes at $r=\infty$.  In a harmonic trap, these are stable or
experimentally stable for sufficiently small nonlinearity. In
contrast to 3D, there is no metastability. We note that, in an
optics context, vortices, or ``spinning bright solitons,'' can also
be stablized by competing nonlinearities, i.e., a defocusing quintic
nonlinearity with a focusing cubic nonlinearity~\cite{towers2001}.

We thank Joachim Brand and William Reinhardt for useful discussions.
LDC thanks the NSF for support. The work of CWC was partially
supported by the Office of Naval Research.


\begin{thebibliography}{10}

\bibitem{saffman1992}
P.~G. Saffman, {\em Vortex Dynamics} (Cambridge Univ. Press, New
York, 1992).

\bibitem{donnelly1991}
R.~J. Donnelly, {\em Quantized Vortices in Helium II} (Cambridge
University
  Press, New York, 1991).

\bibitem{matthews1999}
M.~R. Matthews {\it et~al.}, Phys. Rev. Lett. {\bf 83},  2498
(1999).

\bibitem{madison2000}
K.~W. Madison, F. Chevy, W. Wohlleben, and J. Dalibard, Phys. Rev.
Lett. {\bf
  84},  806  (2000).

\bibitem{shin2004b}
Y. Shin {\it et~al.}, Phys. Rev. Lett. {\bf 93},  160406  (2004).

\bibitem{crasovan2003}
L.-C. Crasovan {\it et~al.}, Phys. Rev. A {\bf 68},  063609  (2003).

\bibitem{qizhou2004}
Q. Zhou and H. Zhai, Phys. Rev. A {\bf 70},  043619  (2004).

\bibitem{anderson2001}
B.~P. Anderson {\it et~al.}, Phys. Rev. Lett. {\bf 86},  2926
(2001).

\bibitem{dalfovo1999}
F. Dalfovo, S. Giorgini, L.~P. Pitaevskii, and S. Stringari, Rev.
Mod. Phys.
  {\bf 71},  463  (1999).

\bibitem{fetter2001}
A.~L. Fetter and A.~A. Svidzinsky, J. Phys.: Condens. Matter {\bf
13},  R135
  (2001).

\bibitem{sulem1999}
C. Sulem and P.~L. Sulem, {\em Nonlinear Schr\"odinger Equations:
Self-focusing
  Instability and Wave Collapse} (Springer-Verlag, New York, 1999).

\bibitem{ruprecht1995}
P.~A. Ruprecht, M.~J. Holland, K. Burnett, and M. Edwards, Phys.
Rev. A {\bf
  51},  4704  (1995).

\bibitem{ueda1998}
M. Ueda and A.~J. Leggett, Phys. Rev. Lett. {\bf 80},  1576  (1998).

\bibitem{sackett1998}
C.~A. Sackett, H.~T.~C. Stoof, and R.~G. Hulet, Phys. Rev. Lett.
{\bf 80},
  2031  (1998).

\bibitem{donley2001}
E.~A. Donley {\it et~al.}, Nature {\bf 412},  295  (2001).

\bibitem{carr2002b}
L. Khaykovich {\it et~al.}, Science {\bf 296},  1290  (2002).

\bibitem{strecker2002}
K.~E. Strecker, G.~B. Partridge, A.~G. Truscott, and R.~G. Hulet,
Nature {\bf
  417},  150  (2002).

\bibitem{caveat}
In fact, the bright solitons investigated in these experiments were
not truly
  in the one-dimensional regime of the mean field~\cite{carr2002c}.
  However, the dynamics and features of interest were all one dimensional.

\bibitem{fischer2004}
U.~R. Fischer, Phys. Rev. Lett. in press  (2004).

\bibitem{chiao1964}
R.~Y. Chiao, E. Garmire, and C.~H. Townes, Phys. Rev. Lett. {\bf
13},  479
  (1964).

\bibitem{moll2003}
K.~D. Moll, A.~L. Gaeta, and G. Fibich, Phys. Rev. Lett. {\bf 90},
203902
  (2003).

\bibitem{firth1997}
W.~J. Firth and D.~V. Skryabin, Phys. Rev. Lett. {\bf 79},  2450
(1997).

\bibitem{carr2004i}
L.~D. Carr and C.~W. Clark, submitted to Phys. Rev. A, e-print
cond-mat/0408460
   (2004).

\bibitem{petrov2000}
D.~S. Petrov, M. Holzmann, and G.~V. Shlyapnikov, Phys. Rev. Lett.
{\bf 84},
  2551  (2000).

\bibitem{press1993}
W.~H. Press, S.~A. Teukolsky, W.~T. Vetterling, and B.~P. Flannery,
{\em
  Numerical Recipes in C: The Art of Scientific Computing} (Cambridge Univ.
  Press, Cambridge, U.K., 1993).

\bibitem{vakhitov1973}
M.~G. Vakhitov and A.~A. Kolokolov, Radiophys. Quantum Electr. {\bf
16},  783
  (1973).

\bibitem{perez1997}
V.~M. P\'erez-Garc\'ia {\it et~al.}, Phys. Rev. A {\bf 56},  1424
(1997).

\bibitem{pu1999}
H. Pu, C.~K. Law, J.~H. Eberly, and N.~P. Bigelow, Phys. Rev. A {\bf
59},  1533
   (1999).

\bibitem{saito2002}
H. Saito and M. Ueda, Phys. Rev. A {\bf 65},  033624  (2002).

\bibitem{dodd1996a}
R.~J. Dodd, J. Research NIST {\bf 101},  545  (1996).

\bibitem{theocharis2003}
G. Theocharis {\it et~al.}, Phys. Rev. Lett. {\bf 90},  120403
(2003).

\bibitem{towers2001}
I. Towers {\it et~al.}, Phys. Lett. A {\bf 288},  292  (2001); R.~L.
Pego and H.~A. Warchall, J. Nonlinear Sci. {\bf 12},  347 (2002).

\bibitem{carr2002c}
L.~D. Carr and Y. Castin, Phys. Rev. A {\bf 66},  063602 (2002);
L.~D. Carr and J. Brand, Phys. Rev. Lett. {\bf 92},  040401 (2004).

\end{thebibliography}

\end{document}